# A superconductor electromechanical oscillator and its potential application in energy storage ( with corrections).

**Osvaldo F Schilling**

Departamento de Física, Universidade Federal de Santa Catarina, Campus, Trindade, 88040-900, Florianópolis, SC. Brazil.

**Abstract.** We discuss theoretically the properties of an electromechanical oscillating system whose operation is based upon the cyclic conservative conversion between gravitational potential, kinetic, and magnetic energies. The system consists of a superconducting coil subjected to a constant external force and to magnetic fields. The coil oscillates and has induced in it a rectified electrical current whose magnitude may reach hundreds of Ampere. The design differs from that of most conventional superconductor machines since the motion is linear (and practically unnoticeable depending on frequency) rather than rotatory, and  it does not involve high speeds. Furthermore, there is no need for an external electrical power source for the system to start out. We also show that the losses for such a system can be made extremely small for certain operational conditions, so that by reaching and keeping resonance the system main application should be in the generation and storage of electromagnetic energy.

The demand for better electrical energy storage technology and cleaner power supplies has increased in recent years. Superconducting magnetic energy storage (SMES) technologies fulfil such requirements[1]. SMES usually involves the establishment of a permanent current in a solenoid which will keep the corresponding magnetic energy for future use. In this paper we discuss a somewhat different design for a SMES machine. Its operation principle is based upon the possibility of cyclic and conservative energy conversion between the lossless electronic motion inside a superconducting coil and the lossless mechanical oscillation associated with levitation of the coil in the presence of a system of magnets. Levitation itself is not obtained by the usual effect of repulsion against a dipolar magnetic field, like in other applications[2], but comes straight from a constant magnetic field through the Lorentz forces it imposes upon the currents in the coil. Electrical currents are obtained from mechanical motion and vice-versa. The motion is linear, rather than rotatory, and high-speeds are not necessarily involved. To achieve such electromechanical motion conversion the superconducting coil is subjected simultaneously to static uniform magnetic fields and to an external force. The imposition of an external static force is fundamental to the method in order to generate mechanical potential energy. For such conditions the coil will oscillate mechanically around an equilibrium position, and will have induced in it a rectified supercurrent containing an alternating component of same frequency. We will discuss the conditions necessary for such system to conserve its initial energy. Hundreds or even thousands of Ampere of rectified current may be obtained and the corresponding energy may either be kept stored in the magnetic field for future use, or immediately transferred for storage or direct utilization elsewhere.

Let´s consider the model experimental setup described in Fig. 1. A rectangular superconducting coil of mass $m$ is submitted to *uniform* magnetic fields $B_1$, $B_2$ and $B_3$ from a system of magnets. It is important that the entire coil be subjected to such fields, as discussed later. The coil is pulled away from the magnet by a constant external force $F$, which in the vertical position might be the coil own weight. According to Faraday's Induction Law the motion of the coil in the presence of magnetic fields will induce in it a transport supercurrent $i$. Such currents will generate different (Lorentz) magnetic forces in the upper and lower sides of the coil. We will assume $B_1 > B_2 > B_{c1}$ ($B_{c1}$ is the superconductor lower critical field). The coil will move with speed $v$ described by Newton´s Law:

$$m \, dv/dt = F - ia(B_1 - B_2) \qquad (1)$$

We introduce the parameter $B_0 \equiv B_1 - B_2$ to simplify the notation, and the length $a$ is the horizontal size of the central magnets – cf. figure. The displacement of the coil gives rise to an induced electromotive force $\varepsilon$, given by Faraday's Induction Law[3] ( for a justification for the utilization of Faraday´s Law in a non-inertial frame of reference see Appendix 1):

$$\varepsilon = -d\Phi_m/dt - L\, di/dt \qquad (2)$$

Here $\Phi_m$ is the magnetic flux from the magnets that penetrates the rectangular area bound by the coil, and $L$ is the self-inductance of the coil. All dissipative processes will be neglected in this theoretical treatment (a hypothesis that will be extensively discussed later), so that the electromotive force in (2) will be considered zero. This results in the conservation of the total magnetic flux ($\Phi_m + Li$) inside the area bound by the coil. In (2), $d\Phi_m/dt = -B_0 a v$. From (2) one obtains a relation between $v$ and $di/dt$. Taking the time derivative of (1) and eliminating $di/dt$ from (2) one obtains:

$$m\, d^2v/dt^2 = -(B_0^2 a^2/L)\, v \qquad (3)$$

This means that the coil shall perform an *ideal* oscillating motion under the action of the external and magnetic forces. Assuming zero initial speed and an initial acceleration equal to $F/m$ eq. (3) can be solved:

$$v(t) = F/(m\Omega)\, \sin(\Omega\, t) \qquad (4)$$

The natural frequency of the oscillations is $\Omega = B_0 a/(mL)^{1/2}$. The amplitude of the oscillating motion is $x_0 = F/(m\Omega^2)$, which may be quite small since it is inversely proportional to $\Omega^2$. It is possible then to combine (1)–(2) to obtain an equation for the current $i(t)$:

$$(B_0 a/\Omega^2)\, d^2 i/dt^2 = F - B_0 a i \qquad (5)$$

whose solution is

$$i(t) = (F/(B_0 a))\,(1 - \cos(\Omega t)) \qquad (6)$$

for $i(0) = di/dt(0) = 0$. From eq. (6) we conclude that the supercurrent induced in the coil is already rectified. It never changes sign, and it looks like the result of submitting an alternating current of frequency $\Omega/2$ to a diode bridge rectifier. Typical figures for the speed amplitude ($v_o$), oscillation amplitude, and rectified current amplitude ($i_o = F/(B_0 a)$) may be obtained. Let's take $L = 10^{-7}$ H, $B_0 = 0.3$ T, $m = 1$ kg, $F = 10$ N, $a = 0.1$ m. For these parameters, $\Omega = 95$ rad/s, $i_o = 333$ A, $v_o = 11$ cm/s, and $x_o = 1.1$ mm. That is, a large rectified current may be obtained with low speed and frequency of oscillation, and very small displacements of the coil.

The set of eq. (1)-(6) was deduced under the assumption of the absence of energy losses. Let's discuss the conditions for such assumption to apply. First of all, it is important to stress that the predicted mechanical motion can be made perfectly frictionless with this design, since in the vertical position the motion is independent of any physical contact between the levitating coil and the magnet. If friction losses are negligible, the next issue that immediately arises is that of the possible losses related to the alternating current in the wire and to inductive coupling between the coil and the magnets. Inductive coupling between the coil and a conducting magnet like Nd-Fe-B will produce resistive eddy currents in the magnet[2]. This source of dissipation might be eliminated by using an insulating magnet like ferrite, at the possible cost of having to work at lower fields. Assuming that this source of energy dissipation may be circumvented let's discuss how to avoid losses associated with the transport of ac currents by the superconductor wire[4-6]. Such losses may be classified in resistive losses and hysteretic losses. Resistive losses might arise due to partial current transportation by normal electrons. The influence of normal electrons may be avoided by working at low frequencies[4,5]. We note that the number of free parameters of the model allows the frequency $\Omega$ to be set, e.g., in the 10 ~ 1000 rad/s range, so that the influence of eddy current losses due to normal carriers expected already in the upper MHz range[5] may be entirely neglected. Alternating currents give rise to a resistive response from the superconducting electrons also[6]. However, well below $T_c$ such resistivity is smaller than the normal state resistivity by a factor $h\Omega/(2\pi\Delta)$, where $\Delta$ is the energy gap[6]. Such factor is of the order of $10^{-9}$ for low frequencies, so that the resistive behavior of superconducting electrons may safely be neglected.

Hysteresis losses occur whenever the flux line (FL) lattice inside a type-II superconductor is cyclically rebuilt by an oscillating magnetic field[4]. In the present case such self-fields ( ripple fields) are created by the alternating part of the current. The work of Campbell, Lowell, and others[7-9] has shown that provided the displacements $d_o$ of the FL from their pinning sites are small enough, such displacements are elastic and reversible, and *no* hysteresis losses occur. The threshold value of $d_0$ was found to be a fraction of the inter-FL spacing, as small as the coherence length ( 2 ~ 6 nm)[7]. Let *b* represent the ripple-field amplitude at the surface of the wire, and *B* the static magnetic field. Campbell[7] has shown that there will be no losses if $b < (\mu_0 B J_c d_0)^{1/2} = b_0$, where $J_c$ is the critical current density and $d_0$ is given above. This explains why the entire coil should be submitted to the static magnetic fields. To avoid self-field hysteresis losses the value of *B* must be large compared to *b*, otherwise the inequality will not be satisfied. The current *i* will flow within a surface sheath of thickness $(B d_0/(\mu_0 J_c))^{1/2}$. There will be no losses associated with this current provided it generates a surface field $\mu_0 i/(2\pi r)$ smaller than $b_0$, and this establishes a criterion for the maximum current attainable in the lossless vortex motion regime[10]. This oscillator is probably the first superconductor device in which the transport of currents in the regime of reversible vortex oscillations finds a technical application. Just to give a numerical example, if a cold-drawn Nb-Ti wire of radius $r = 0.085$ mm, and $J_c = 5 \times 10^9$ A/m$^2$[11] is used to make the coil, there will be no hysteresis losses for a transport current as high as 1.5 A in such thin wire ( adopting $B = 1$ T). Of course, much higher currents are possible if thicker conductors are used, even in multifilamentary form.

This overall discussion has therefore shown how hysteresis losses may be avoided in the design of this superconducting machine. Associated also with FL motion, resistivity losses originated from the motion of the normal electrons in the cores of the FL ( flux-flow resistivity) can be decreased to a factor of about $10^{-8}$ times the normal state resistivity by working at low frequencies and high $J_c$ materials. One might mention also flux-creep as another possible source of dissipation, but the use of a temperature of operation well below $T_c$ , and a strong-pinning material should discard the possibility of creep.

We have discussed the major possible sources of dissipation and how to avoid them. It is possible to conclude that losses may be severely restricted, if not fully eliminated, by a proper choice of magnet design and magnetic material together with a proper design of the coil and operation conditions.

Nevertheless, it seems important also to quantitatively estimate the effects of external loads upon the behavior of the system. Let's consider the effect of adding a resistance $R$ to the circuit. In this case equation (2) becomes:

$$0 = d\Phi_m/dt + L\, di/dt + R\, i \qquad (7)$$

Following the same procedure as before it is possible to calculate both current and speed as a function of time by solving eqs. (1) and (7). The speed ceases to be purely oscillating, displaying an additional component that is initially a linear function of time and converges to a constant limit at long times. Of course this is a consequence of energy losses in the resistor, which will make the coil progressively escape from inside the magnets region under the effect of $F$. However, the oscillating component of the speed will be dominant at t=0+ provided $R/L < 2\,\Omega$. The current produced is rectified, displaying damped oscillations, whose frequency becomes $\Omega_R = (\Omega^2 - R^2/(4L^2))^{1/2}$. The frequency must be a real number and this imposes a limit on the maximum stray resistance that the system may sustain. For the numerical example given before $R$ should be smaller than $2\times10^{-5}$ ohm, but preferably much smaller than this. It must be pointed out, however, that the utilization as a SMES does not actually depends on the oscillator being capable of working long times with little or no dissipation. The coil might simply be hung at rest and put to oscillate only when energy is actually needed. If the condition $R/L < 2\Omega$ is satisfied this will guarantee that the maximum current, and thus maximum stored energy, will be obtained at once when oscillations begin and that such energy will immediately be available for transfer or direct external use. This leads us to an important application of this system, which may be obtained by adding a capacitor $C$ to the circuit. The oscillator will then behave like a generator, transferring charge and energy to the capacitor. Equation (2) becomes:

$$0 = d\Phi_m/dt + L\, di/dt + q/C \tag{8}$$

The current produced is rectified, with *undamped* oscillations of frequency $\Omega_C = (\Omega^2 + 1/(LC))^{1/2}$. The capacitor will continuously be loaded, so that the time-averaged potential difference $V$ between its plates increases at a rate $dV/dt = (F/(B_0 aC))(\Omega/\Omega_C)^2$. Again the speed of the coil displays a component that varies linearly with time, but such speed is inversely proportional to $C$ for very large values of $C$. This indicates that the oscillator might operate connected to a large capacitance bank with only minor alterations upon its unloaded behavior. Again, an important operational detail is that as long as the coil oscillates inside the magnets region there will be a *maximum* current flowing. Therefore, it seems recommendable that the system includes a device that will reverse the direction of the force $F$ at fixed time intervals. At each reversal the system is restarted. This will keep the system in resonance conditions and will avoid the escape of the coil in case of losses, allowing the system to operate in permanent mode.

In conclusion, the present work has theoretically described a superconducting electromechanical oscillator in which losses might practically be eliminated, providing a technique for the direct conversion of gravitational energy into electromagnetic energy. The machine described here might be used as a "stand alone" SMES machine and generator, which would generate, store, and transfer magnetic energy to ( for instance) a bank of capacitors, with no need of an external power supply for it to start out. We have extensively discussed the conditions required to avoid dissipative effects due to normal resistivity, hysteresis, and flux-creep, so that the system may operate in a permanent mode. The actual implementation of a system with these characteristics seems perfectly possible in view of the available materials and magnet fabrication technologies.

The author wishes to thank Prof. Said Salem Sugui Jr. for his support. The author is grateful also to Professor A.M. Campbell for his comments and for clarifying details of his work on the reversible movement of flux lines.

APPENDIX 1: Corrections to Maxwell Equations in a non-inertial frame of reference.

The model discussed in this work is based on the validity of the usual form of Faraday's Induction Law, equation (2), in the frame of reference of the oscillating coil. However, such frame is non-inertial since the coil is accelerated by the combined effects of the external force *F* and the Lorentz forces. In order to put the basic equations of Electromagnetism in a form valid for a non-inertial frame of reference it is necessary to replace all partial derivatives by covariant derivatives, and calculate the effect of the acceleration upon the metric tensor $g_{ik}$ that characterizes the space[12], utilizing for this purpose the same mathematical tools employed in Gravitation Theory. In this Appendix we will show that the corrections needed in Maxwell's equations are negligible in this case, so that equation (2) may be accepted as correct even in the oscillating coil frame. From the formal mathematical point of view the corrections to the Maxwell equations for non-inertial frames of reference are contained in the quantities $(-G)^{1/2}$ and $g_{ik}$, where *G* is the determinant of the metric tensor $g_{ik}$[12]. It is well known that for strictly inertial frames of reference the space-time interval obeys the equation (ref. 12, section 82):

$$-ds^2 = -c^2 dt^2 + dx^2 + dy^2 + dz^2 = g_{ik} dx^i dx^k \tag{A1}$$

where repeated indexes are summed from 0 to 3. In this case the components of the metric tensor $g_{ik}$ are: $g_{11}=g_{22}=g_{33}=1$, $g_{00}=-1$, the $g_{ik}=0$ for i≠k, $G=-1$, and the Maxwell equations are valid with their usual expressions. Let's assume that the solution to the problem developed in the article is correct and then obtain expressions for $G$ and for the $g_{ik}$ to be compared with the corresponding values for strictly inertial frames of reference given above. In Figure 1 let $x$ represent the vertical coordinate measured by an inertial frame fixed to the magnet, and $x'$ the coordinate measured by a frame fixed to the coil. The two coordinate systems are related by the equations ( neglecting further relativistic corrections in view of the very low speed of the coil compared to $c$): $x = x' + x_o \sin(\Omega t)$, $y = y'$, $z = z'$, $t = t'$. Here $x_o$ is the amplitude of the oscillating motion and $\Omega$ is the natural frequency of the oscillations calculated in the article. Calculating the derivative of $x$ one obtains:

$$dx = dx' + \Omega x_o \cos(\Omega t)\, dt \qquad (A2)$$

In the primed frame of reference the space-time interval is:

$$-ds^2 = \{(\Omega^2 x_o^2/c^2) \cos^2(\Omega t) - 1\} c^2 dt^2 + 2\Omega x_o \cos(\Omega t)\, dx'\, dt + dx'^2 + dy'^2 + dz'^2 \qquad (A3)$$

One notices at once that $g_{00}' = -1 + (\Omega^2 x_o^2/c^2)\cos^2(\Omega t)$, and that the only non-vanishing non-diagonal coefficients are $g_{01}' = g_{10}' = (\Omega x_o/c)\cos(\Omega t)$. This makes the determinant $G' = -1$, which is the same as $G$ in the inertial frame of reference. Taking the figures obtained in the numerical example given in the article, i.e., $x_o = 1.1$ mm, $\Omega = 95$ rad/s, the following values for the altered coefficients are obtained ( the remaining coefficients are exactly the same as those for the inertial frame of reference):

$g_{00}' \approx -1 + \Omega^2 x_o^2/c^2 = -1 + 10^{-19}$.

$g_{01}' \approx \Omega x_o/c = 3 \times 10^{-10}$.

In conclusion, what these calculations have shown is that for the specific problem of a superconducting coil oscillating at low frequency and amplitude, the Maxwell equations in general, and Faraday's Law (equation (2)) in particular, may be used with their usual expressions valid for inertial frames of reference with a precision of about 1 part in $10^{10}$.

Figure 1: A rectangular superconducting coil is submitted simultaneously to an external force *F* and to magnetic fields $B_1$, $B_2$ and $B_3$ (greater than $B_{c1}$) perpendicular to it. As shown in the text the predicted motion is oscillatory, with a rectified current being induced in the coil. Note that during oscillations all sections of wire are kept under constant fields in order to decrease or eliminate hysteresis losses.

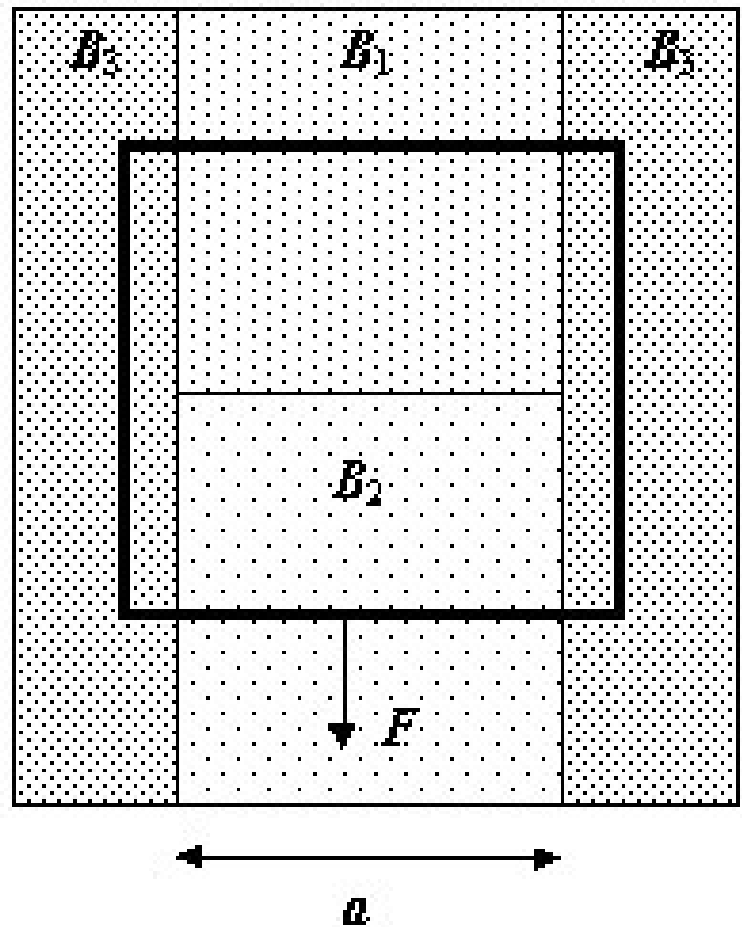